\documentclass{article}
\usepackage{graphicx}
\usepackage{amssymb}

\begin{document}
\begin{titlepage}
\begin{center}
\Large
{\bf Comparison of several tetrahedra-based lattices}

\vspace{5mm}
\normalsize
Maged~Elhajal, Benjamin~Canals, Claudine~Lacroix

\vspace{5mm}
\small
Laboratoire Louis N\'eel, 25 avenue des Martyrs, 
BP 166, 38042 Grenoble Cedex 9, France

\vspace{5mm}
June 2000

\vspace{5mm}
\normalsize

A comparison of the quantum Heisenberg antiferromagnetic model
(QHAM)
on the pyrochlore lattice, the checkerboard lattice and the square lattice with crossing 
interactions is performed. 
The three lattices 
are constructed with the same tetrahedral unit cell and this property 
 is used to 
describe the low energy spectrum by means of an effective
hamiltonian restricted to the singlet sector. We analyze the structure of the 
effective hamiltonian and solve it within mean 
field approximation for the three lattices.

\end{center}
\end{titlepage}


\section{Introduction}

Geometrical frustration gives rise to original magnetic properties at low temperature: on some lattices, a continuous degeneracy of the classical ground state is obtained in the mean field approximation, preventing any phase transition to a long range ordered state\cite{ReimersBerlinskyShi}. 
We will focus here on the quantum spin-$\frac{1}{2}$ Heisenberg model on the 
checkerboard, the pyrochlore and the square lattice with crossing 
interactions and specially on their low energy spectrum.

\begin{figure}
\begin{center}
\includegraphics[width=8cm]{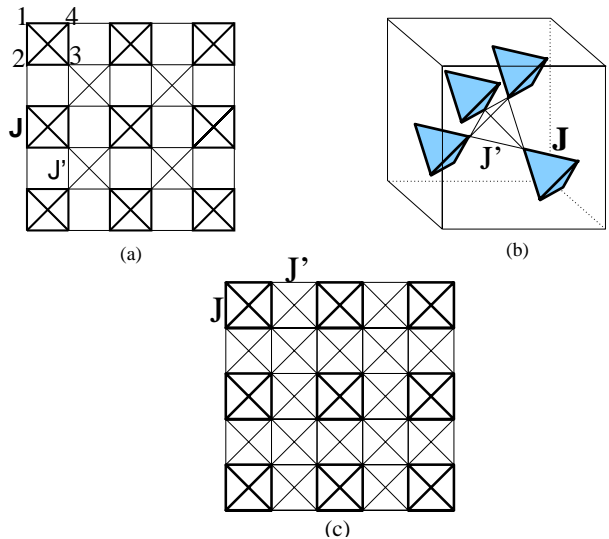}
\caption{\it Checkerboard (a), pyrochlore (b) and square lattice with crossing interactions. Bold lines represent intra-TT (J) interactions treated exactly, and thin lines represent inter-TT (J') interactions considered as a perturbation.}
\label{figure1}
\end{center}
\end{figure}

The checkerboard lattice is a square lattice of tetrahedra (TT) (see figure \ref{figure1}a), while the pyrochlore lattice is a fcc lattice of TT (see figure \ref{figure1}b).
The checkerboard can be considered as the 2D analog of the pyrochlore since both  of them are made of corner sharing tetrahedra. The square lattice with 
crossing interactions is also a 2D lattice but the TT are more connected than 
in the checkerboard (see figure \ref{figure1}c).

There is strong evidence, both experimentally\cite{BallouLelievre-BernaFak} and theoretically 
\cite{BenjaminClaudine} that the ground state for the QHAM on the pyrochlore lattice is a 
quantum spin liquid, with a correlation length which hardly exceeds a few interatomic distances. Moreover, it is believed that the low lying excitations are singlet-singlet as in the kagome lattice\cite{LecheminantBernuLhuillierPierreSindzingre}. Given these properties, we develop an effective model restricted to the singlet sector of the Hilbert space and taking into account the short range correlations of the spin liquid state. For the checkerboard, it was 
shown\cite{LiebSchupp} that the ground state is a singlet state. The square lattice 
with crossing interactions is similar to the $J_1-J_2$ model with $J_1=J_2$. The ground 
state for this model is not clearly identified\cite{SinghWeihongHamerOitmaa} for 
$0.4\lesssim\frac{J_2}{J_1}\lesssim0.6$ but in our case ($J_1=J_2$), the ground state is 
 a columnar dimer state\cite{KotovZhitomirskySushkov}.


\section{Effective hamiltonians}

The ground state of one TT is twofold degenerate (2 singlet states). 
We restrict the accessible states for each TT to these two singlets, and associate a pseudo-spin $\stackrel{\rightarrow}{T}$ (with T=$\frac{1}{2}$) to each TT, the two states ($T^z=\pm \frac{1}{2}$) of the pseudo-spin represention being the two singlet states. 
Doing this, we have only taken into account (exactly) the interactions (J) within one TT. 
We label the 4 sites of each TT as shown in figure \ref{figure1}a. 
We can  choose the two states of the S=0 subspace so that 

\makebox{\includegraphics{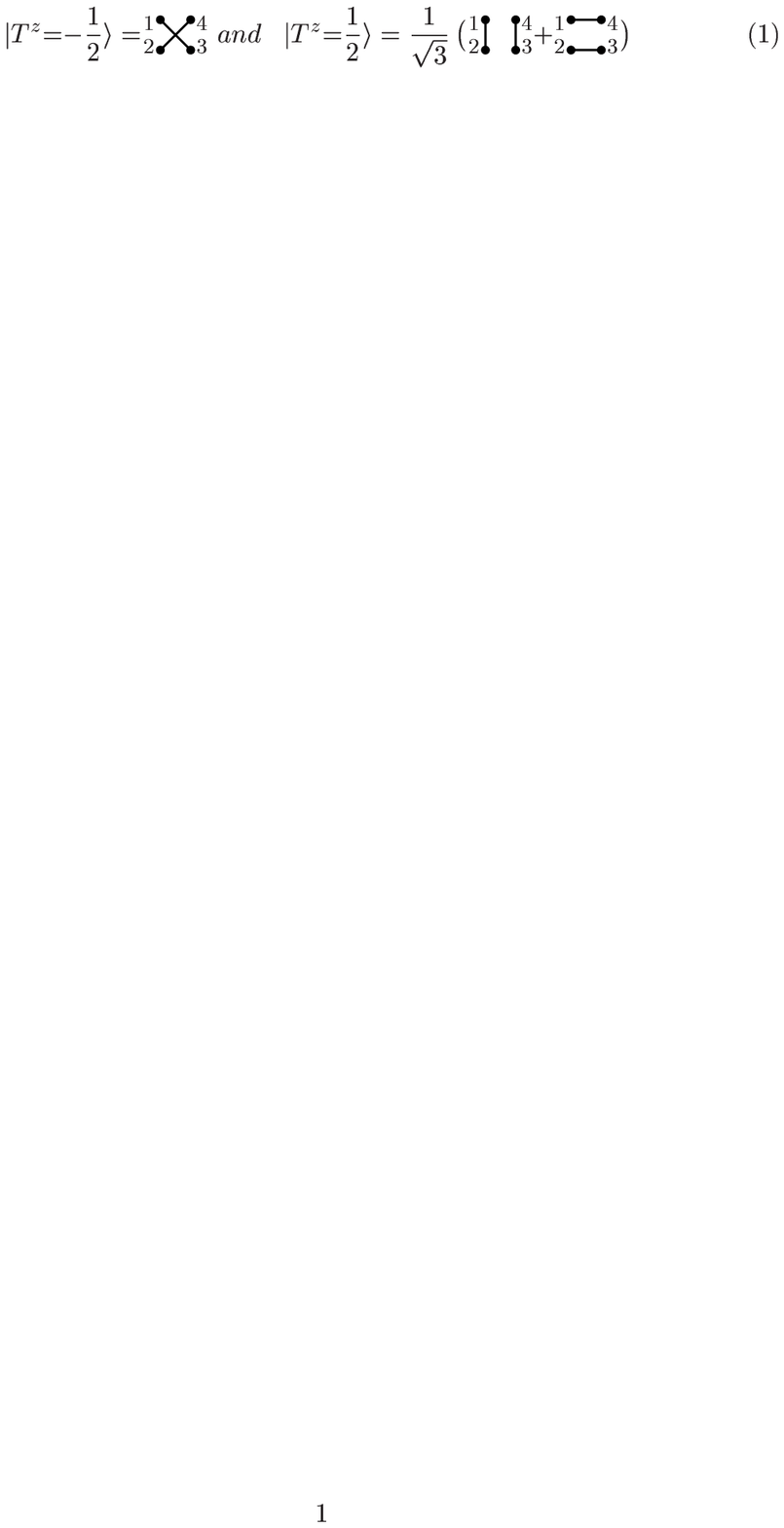}}

where 
\makebox{\includegraphics{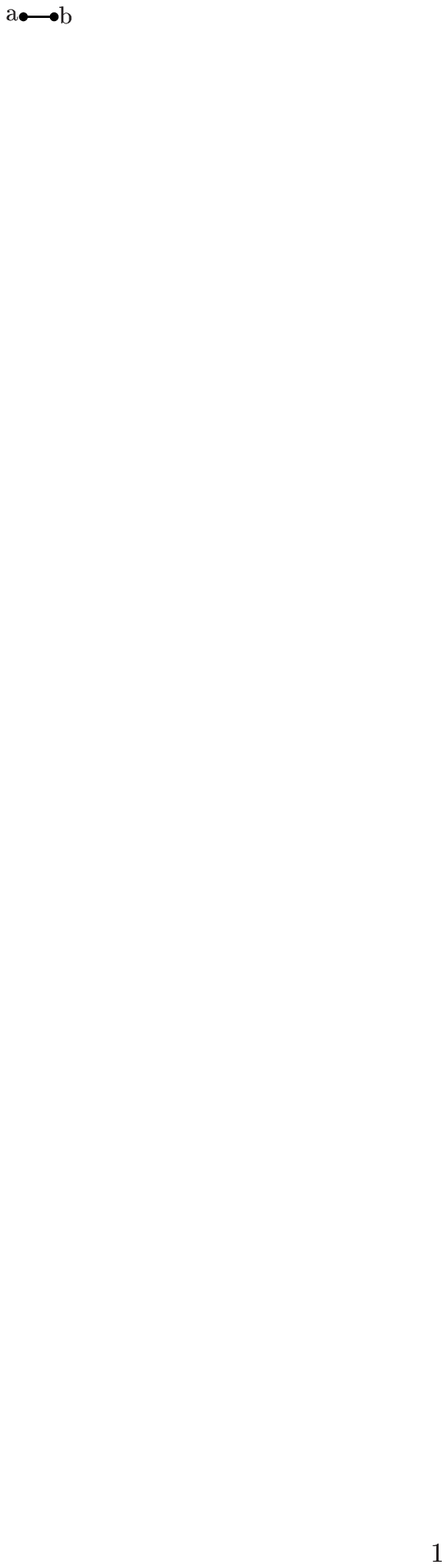}}
represents a 2 spins singlet $\frac{1}{\sqrt{2}}\left(\mid a\uparrow b\downarrow\rangle-\mid a\downarrow b\uparrow\rangle\right)$. $\mid\hspace{-1mm} T^{z}\hspace{-1mm}=\hspace{-1mm}-\frac{1}{2}\rangle$ describes a dimerized state whereas $\mid\hspace{-1mm} T^{z}\hspace{-1mm}=\hspace{-1mm}\frac{1}{2}\rangle$ is a linear combination of dimerized states, which is a quadrimer.
We then consider the interactions between the tetrahedra (J') as a perturbation. 
Since the two singlets of one TT are degenerate, perturbation theory leads to a matrix which we write down in terms of pseudo-spins operators ($T^x$, $T^y$ and $T^z$). 
This leads to an effective hamiltonian in pseudo-spins space.
These hamiltonians have been calculated up to third order in 
$\lambda=\frac{J'}{J}$ for all lattices. For the three lattices at first order, there is no 
effective interaction (the effective hamiltonian is a spherical matrix) because the operators in real spin space ${\bf s_i}{\bf s_j}$ have non-zero matrix elements only between singlet and triplet states.

\begin{figure}
\begin{center}
\includegraphics[width=2cm]{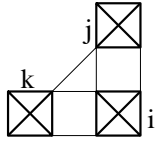}
\caption{\it Three pseudo-spins interactions in the checkerboard include all possible triangles like this one and those obtained by reflexion with respect to the x and y axes, keeping always the same labelling (i, j, k) for the pseudo-spins (see fifth sum of the effective hamiltonian)}
\label{figure2}
\end{center}
\end{figure}

For the checkerboard lattice, we find (with $J'=J$ and summing the contributions up to third order): 

\begin{eqnarray*}
\frac{H}{\vert J\vert}&=&\sum_{psd-spins} -\frac{85}{192}-\frac{1}{24}T^z_i\\
&+&\sum_{horiz} \frac{1}{16}T^z_i T^z_j -\frac{1}{16}T^x_i T^x_j-\frac{1}{16\sqrt{3}}\left(T^z_iT^x_j +T^x_iT^z_j\right)\\
&+&\sum_{verti} \frac{1}{16}T^z_i T^z_j -\frac{1}{16}T^x_i T^x_j+\frac{1}{16\sqrt{3}}\left(T^z_iT^x_j +T^x_iT^z_j\right)\\
&+&\sum_{NNN} -\frac{1}{48}T^z_i T^z_j +\frac{1}{16}T^x_i T^x_j\\
&+&\sum_{triangles}\frac{1}{24}T^z_iT^z_jT^z_k+\frac{1}{8\sqrt{3}}\left(T^z_iT^z_jT^x_k-T^z_iT^x_jT^z_k\right)\\
&&\hspace{1cm}-\frac{1}{8}T^z_iT^x_jT^x_k
\end{eqnarray*}

where the first sum is over pseudo-spins, which are on a square lattice for 
the checkerboard.
The second and third  sums are over horizontal (x axis) and vertical (y axis) nearest neighbour 
interactions, whereas the fourth sum is over second nearest neighbours. 
The fifth sum is over all possible triangles and the pseudo-spins are then 
labelled as shown in figure \ref{figure2}.

The 'field' acting on pseudo-spins has nothing to do with a real magnetic field and it is hopeless to measure any 'real' magnetization, since all states in our representation are non magnetic.
The interactions along $x$ and $y$ directions of the square lattice are 
slightly different. 
This broken symmetry has been introduced artificially when we have choosen 
two arbitrary singlets for each \hyphenation{TT} TT which are not invariant under rotation of 
$\frac{\pi}{2}$, but the final result does not depend on this choice. 
The second and third sums lead to anisotropic exchange and to cross terms.  
There are three pseudo-spins interactions at third 
order in $\lambda$.
The higher the order of perturbation is, the higher the number of spins involved in the interactions will be. 
However, since the correlation length is short, it is believed that taking into account the lowest orders in perturbation will be enough to catch the essential feature of the spin liquid behaviour. 

The 'field' in pseudo-spins space can be interpreted as follows: since the state $\mid\hspace{-1mm} T^{z}\hspace{-1mm}=\hspace{-1mm}-\frac{1}{2}\rangle$ corresponds to a dimerized singlet 
 and the state $\mid\hspace{-1mm} T^{z}\hspace{-1mm}=\hspace{-1mm}\frac{1}{2}\rangle$ to a quadrimer (see (1)), a negative 'field' would be a hint that the system has a tendency to dimerize, 
whereas a positive 'field' would be a tendency of the system to be in a 4-spin-singlet (quadrimerization). 
This is confirmed by the expression of the effective hamiltonian for the 
square lattice with crossing interactions (at third order in $\lambda$) which 
is known to dimerize (we have set $J'=J$):

\begin{eqnarray*}
\frac{H}{\vert J\vert}&=&\sum_{psd-spins} -\frac{125}{192}+\frac{5}{6}T^z_i\\
&&+\sum_{horiz} -\frac{5}{24}T^z_i T^z_j -\frac{5}{8}T^x_i T^x_j-\frac{5}{8\sqrt{3}}\left(T^z_iT^x_j +T^x_iT^z_j\right)\\
&&+\sum_{verti} -\frac{5}{24}T^z_i T^z_j -\frac{5}{8}T^x_i T^x_j+\frac{5}{8\sqrt{3}}\left(T^z_iT^x_j +T^x_iT^z_j\right).
\end{eqnarray*}
This hamiltonian was calculated at second order by Kotov, Zhitomirsky and Sushkov\cite{KotovZhitomirskySushkov}. It is found 
that the hamiltonian at third order is equivalent  to the hamiltonian 
at second order multiplied by a constant ($\frac{1}{4}$), so including the third order will not change the structure of the hamiltonian. It is similar to the 
hamiltonian for the  
checkerboard at second order except that the field is  
higher and of opposite sign. 
In particular, the ratio of the field term on the interactions terms remains constant when including third order, which reflects the tendency of this system to dimerize (because the 'field' is strong enough and negative, so that the pseudo-spins will rather be in the state $\mid\hspace{-1mm} T^{z}\hspace{-1mm}=\hspace{-1mm}-\frac{1}{2}\rangle$ which is a dimer). On the contrary, in 
the checkerboard case, inclusion of third order perturbation decreases the ratio between 'field' and exchange. 
Whether this is still true at fourth order is being tested.
 The sign of the field in the checkerboard is a hint of quadrimerization

The hamiltonian for the pyrochlore has already been published in \cite{HarrisBerlinskyBruder}. 
Our expression of the effective hamiltonian is slightly different (the constant term is different).
The main difference with the effective hamiltonian for the checkerboard is the absence of 'field' up to third order in $\frac{J'}{J}$, and the absence of interactions at second order. The three 
pseudo-spins interactions are the same as for the checkerboard. Labelling 
three pseudo-spins interactions is more complex than for the checkerboard case 
because the pseudo-spins are on a 3D (fcc) lattice. 


\section{Mean field approximation}

We have solved the effective hamiltonians for the three lattices within mean field approximation. 
We have first looked for single-{\bf q} solutions, taking into account only 
two pseudo-spins interactions. 
For the pyrochlore and the checkerboard lattices, it is found that the 
structure of the ground state is antiferromagnetic in pseudo-spin space. 

For 
the checkerboard lattice, the propagation wave vector is ($\pi$, $\pi$). 
The two sublattices have a pseudo-magnetization which is along $\pm z$.  
Going back to real spin space, this means that on one sublattice the TT are 
in a dimerized state (corresponding to pseudo-spin $\mid\hspace{-1mm} 
T^{z}\hspace{-1mm}=\hspace{-1mm}-\frac{1}{2}\rangle$), whereas the TT on the 
other sublattice are quadrimers (in pseudo-spin space: $\mid\hspace{-1mm} 
T^{z}\hspace{-1mm}=\hspace{-1mm}\frac{1}{2}\rangle$). 

For the square lattice with crossing interactions, all the TT are in the 
$\mid\hspace{-1mm} T^{z}\hspace{-1mm}=\hspace{-1mm}-\frac{1}{2}\rangle$ state 
at zero temperature, which correspond to a fully dimerized state, as expected\cite{KotovZhitomirskySushkov}.

For the pyrochlore lattice, the wave vector can be either ($2\pi$,0,0), (0,$2\pi$,0) or (0,0,$2\pi$). The structure is still antiferromagnetic (one dimerized and one quadrimerized sublattices), but the direction of the pseudo-magnetization depends on the wave vector.  Each wave vector corresponds to one of the 
 three possible dimerized states for 
a TT: 
\raisebox{-2mm}{\makebox{\includegraphics{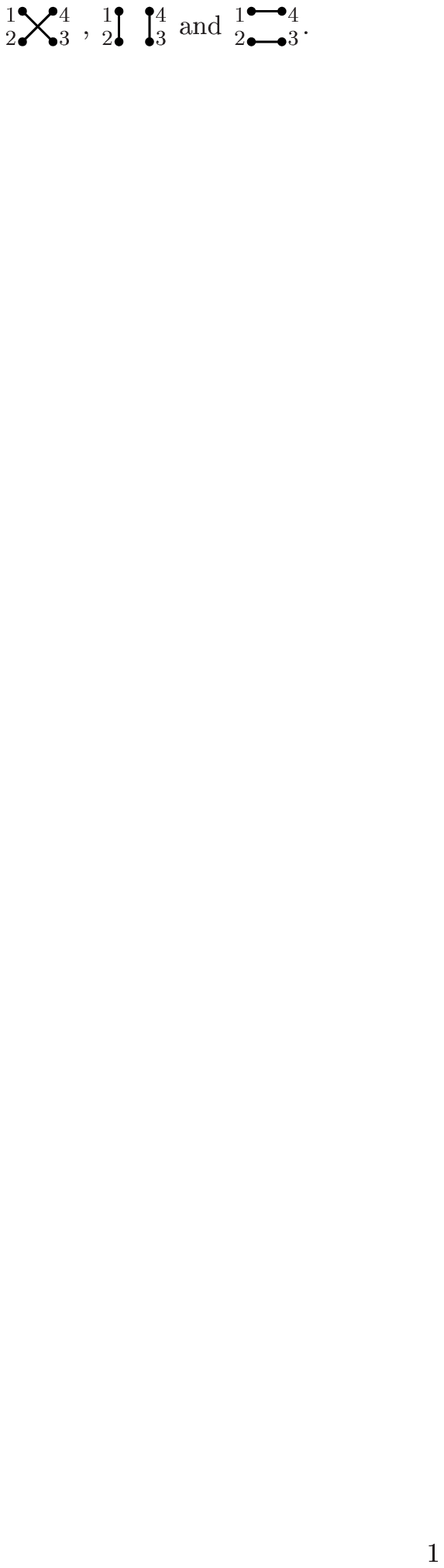}}}
In these antiferromagnetic single-{\bf q} configurations, the contribution of 
the three pseudo-spins interactions vanishes for both lattices. 

For the pyrochlore lattice we have then looked for multi-{\bf q} structures as done in Ref. \cite{HarrisBerlinskyBruder}. 
A four sublattices stucture is found to be 
favoured by three pseudo-spins interactions. It corresponds to a 
triple-{\bf q} structure which is a superposition of the three single-{\bf q} 
solutions. Three of the sublattices are in 
the three possible dimerized states of a TT, and the 
pseudo-magnetization of the fourth one is: $\langle T^x_i\rangle =\langle T^z_i\rangle =0$ and $\langle T^y_i\rangle =\pm\frac{1}{2}$. The sign of $\langle T^y_i\rangle$ on each site 
is not fixed since the energy is independent of $T^y$ and this sublattice is 
disordered. 

Separating the contributions of the different orders of perturbation, 
the total energy per TT (or pseudo-spin) for this triple-{\bf q} solution is:

$$
E=-\frac{3}{2}|J|-\frac{9}{16}\frac{J'^2}{\mid J\mid}+\frac{9}{128}\frac{\mid J'^3\mid}{J^2}
\label{eqn2}
$$

The modulus of the different energy contributions decrease as the order of perturbation increases, as expected.
Fixing $\frac{J'}{J}=1$, this gives an energy per site ($\frac{E}{4}=-\frac{255}{512} |J|\simeq -0.49805 |J|$) slightly lower than the value obtained in 
Ref. \cite{HarrisBerlinskyBruder} 
($\frac{E}{4}\simeq -0.48763 |J|$).
The energy per site obtained for the checkerboard lattice is $\frac{E}{4}\simeq -0.49622\vert J\vert$

To summarize, our results indicate that the effective hamiltonians are different 
for the pyrochlore, the checkerboard and the square lattice with crossing 
interactions, due to the different geometries. The 'field' in the effective 
hamiltonian is associated to dimerization or quadrimerization. The energies obtained for the ground states in the pyrochlore lattice (3D) and the checkerboard lattice (2D) are quite similar. However, the singlet 
structures of the two lattices are different: the 
pyrochlore lattice has a four sublattices structure, one of which does not 
order, the three other being in dimerized states. The square lattice with crossing interactions has a 
clearly different behaviour at low temperature. Further studies are 
necessary to see whether the structures obtained persist if fluctuations around mean field solution and higher order terms are taken into account.

\begin{center}
{\bf \large acknowledgments}
\end{center}
It is a pleasure to acknowledge valuable discussions with Pr. A.B. Harris and Pr. A.J. Berlinsky.

\end{document}